Ultrahigh permittivity in core-shell ferroelectric ceramics: Theoretical approach and practical conclusions

M. Anoufa[1,2], J. M. Kiat[1,3] and C. Bogicevic[1]

[1]Laboratoire Structures, Propriétés et Modélisation des Solides, CentraleSupelec, CNRS-UMR8580, Grande Voie des Vignes, 92295 Châtenay-Malabry Cedex, France

[2]Nextmat, CentraleSupelec, Grande Voie des Vignes, 92295 Châtenay-Malabry Cedex, France

[3]Laboratoire Léon Brillouin, CE Saclay CNRS-UMR12, 91991 Gif-Sur-Yvette Cedex, France

Abstract:

Ferroelectric anomalies in most classical ferroelectric materials such as BaTiO3 (BT) and related compounds tend to become diffuse and eventually vanish when the grain size decreases. This has led to question the very existence of ferroelectricity for very small particles. However there are also reports in the literature of ultra-high values of permittivity when specific processes or doping are performed, for instance in the BT-family, or in the $CaCu_3Ti_4O_{12}$-family and interpretations using (IBLC) have been proposed. We have considered these effects in the framework of a core-shell model of ceramics. The ultrahigh values mentioned above could be reproduced by considering that the core (be it dielectric or ferroelectric) have non-zero conductivity. In particular a huge background of permittivity and a Maxwell Wagner relaxation were obtained in agreement with the experimental situation. Different types of inhomogeneity (grain size distribution, mixing of conductive and non-conductive phases, porosity) have also been considered and their effects are discussed. Finally we have evaluated the potentiality of energy storage in IBLC compounds and observed a reduction (compared with non-conductive ferroelectrics) by a factor $\approx$2-4 depending on the structuration of the ceramic, which shows that the performance of such compounds are lower than in optimal insulating core-shell structures.

I. INTRODUCTION

Ferroelectric compounds have numerous applications in microelectronics, in particular as ceramics for capacitors and actuators devices but they require a precise control of the size, shape and structuration of the grains. Obviously, sub-micronic grains size is needed to increase the integration of ceramics in systems, but in many cases this has the drawback to alter the performances of the material.

From a general point of view the permittivity ($\epsilon$) of a material results both from intrinsic and extrinsic properties, the latter been associated with possible existence of domains, defects or doping, or with the shape and geometry of the grains, the structuration of the ceramic, existence of internal strain etc etc. Classically, in most usual materials such as $BaTiO_3$ (BT) and related compounds, a diminishing in grain sizes makes the ferroelectric anomalies to become diffuse and eventually to vanish in such a way that the existence of ferroelectricity at very small sizes has been often questioned.

However, in opposition with these reports, some other publications (e.g. ref.1-3) have mentioned instead of a reduction a strong increase of permittivity associated to specific processes such as spark plasma sintering (SPS). In some cases the dependence of $\epsilon$ appears to be completely flat over a wide range of temperature, but generally, residual peaks associated with phase transitions are still observed. For instance in La-doped BT or in $BaTiO_{3-x}$, gigantic values of $10^6$ were reported[2] in nanopowders sintered by SPS, but with strong static loss associated to conductivity. In these experiments it seems that a defective grains surface is at play as vacancies or doping ions do not penetrate homogeneously inside the grain. Also in artificial core-shell structures obtained by coating ferroelectric grains with $SiO_2$, $Al_2O_3$ or others, (written as $BT@SiO_2$ for instance) values of $\epsilon$ superior to $10^5$ has been demonstrated[3]. Figure 1 depicts this type of observation.

This effect is actually used in industrial type III or X7R capacitors[4-5] i.e. when a flat temperature-dependence of permittivity and high capacity are needed, but with the drawback of weak voltage operating range and high frequency dependence. In the case of values as high as those displayed in figure 1 an extrinsic effect seems to superpose on ferroelectricity: indeed, a very flat component seems to be added to the primitive ferroelectric anomalies.

In addition to these observations new materials with very high $\epsilon$ values up to $10^6$ have been discovered in the $ACu_3Ti_4O_{12}$ perovskite family (A = Ca, Ba, Sr, see e.g. ref. 5) but with no ferroelectric-like anomaly (Fig. 2 from Ref. 6 and 7). At low temperatures (i.e. $\approx 200K$) $\epsilon$ drastically decreases and exhibit an Arrhenius-type relaxation with a thermal activation close to 0.1-0.4 eV. This flat component of the temperature-dependence of the permittivity with very high value has been explained as an extrinsic effect associated to the Maxwell-Wagner (MW) relaxation induced by the formation of internal barriers layer capacitors, the so-called IBLC effect. The basic idea[8] is that a core-shell structure is again at play where the core behaves more like a conductor or a semiconductor rather than an insulator while the shell remains a true insulator, which allows charge accumulation at interfaces. In view of understanding the chemical origin of this conductivity, impedance spectroscopy experiments were performed[8] and shown that the internal resistance of the grains is lowered when sintering is performed in nitrogen atmosphere. Therefore, in standard atmosphere conditions, oxygen reduction during sintering probably induces vacancies, which increases the conductivity of the grains. This effect has also been proposed to explain the high permittivity of core-shell BT based materials.

In order to overcome the high static loss and the weak rigidity, special core-shell structuration with specific coating of the grains have been shown to be efficient and patents have been filed. If such solutions would appear effective, electrostatic energy storage at low prices, quasi infinite cyclings, ultrafast charge-discharge time, high power would be feasible whereas up to now energy storage in ferroelectric compounds have been demonstrated to remain only fair[9].

The aim of this paper is to discuss from a global point of view all the effects mentioned above on the basis of the core-shell structuration of the ceramics, in order to explain the different types of thermal dependence of $\epsilon$ that have been reported, the associated diffusivity, the anomalies at critical temperatures, which in some cases are still observed, and to discuss the different parameters at play in these observations. We believe this will help to clarify some confusions in the literature. In particular the high value of permittivity of IBLC is often proposed to have potential applications, whereas a weak electric field destroys these performances because of the conductivity. We will discuss whether there is or there is no practical interest for such effect.

We will use for this purpose as a starting point the electrostatic modeling of core-shell ceramics we have recently developed in the framework of the Landau theory and applied to energy storage[9] and electrocaloric performances[10] of the most usual ferroelectric compounds. This modeling has been compared with effective Hamiltonian approach we also developed for ceramics[11] and was shown to be powerful for the discussion of macroscopic properties. Among the results it was demonstrated that there exists some optima in the core-shell structuration of the ceramics, which give much better performances than in standard ceramic[9]; also playing on the structuration enables to tailor the temperature of the maximal effect so as to place it at an optimal temperature, for instance close to 300K[10]. This model has shown that the core-shell picture of nanoparticles adequately explains experimental data in BT, in particular the size evolution of the critical temperatures: neither surface tension nor strain was needed at a primary stage to describe experimental results. Of course, elastic interactions, surface tension and inhomogeneous strain can have a strong influence but not so much for dielectric properties. The experimental situation in $SrTiO_3$ (STO) core-shell ceramics is a good illustration of these points: the size dependence of the permittivity is explained on the basis of a dielectric model based on a core-shell structure[12] without any strain but the recent evidences of a new (orthorhombic) phase at low temperature for STO@$ZrO_2$ requires a model taking also into account elastic properties [13].

However as we are going to show, this model cannot explain the ultrahigh values of permittivity discussed here. For that purpose we must modify and extend the core-shell picture.

This paper is organized as follow. In section II we recall briefly the basic assumptions and results of the model. Then in section III we apply the model to study the effects of a "perfect" core-shell structure of a ferroelectric ceramic on the temperature-dependence of the permittivity. Then in section IV we consider the resulting modifications induced in the case of a real ceramic with a grain size distribution. Then the following part of the paper is dedicated to the explanation of the IBLC effect. For that purpose we introduce in section V conductivity in the core. We treat first in section V-I the case of non-intrinsic ferroelectrics such as CCTO and then in section V-II the case of ferroelectrics such as BT, first in the case of a perfect core-shell structure and then in the more real case of a size distribution. As porosity is often mentioned as a parasitic effect inducing degradation of performances in ferroelectrics we briefly discuss in section VI this point using the ideas introduced in the preceding sections. Then in section VII we discuss the potential interest of IBLC in the framework of energy storage. Finally practical conclusions are summarized.

II. Analytical model of core-shell particles.

In this section, we briefly recall the model for <u>standard</u> dielectric and ferroelectric materials; for more details, the reader can refer to the dedicated paper (Ref. 9-11). The calculation starts by considering a single core-shell spherical particle built on a ferroelectric core coated by a linear dielectric placed in void and solving the Laplace equation ΔV=0. Note that the calculation has been also performed in the case of anisotropic shapes of the grain such as nanorods or nanoplates, but as the main physical consequences are not changed for the problems that we are dealing in this paper, we will discuss only the spherical case. The two sole parameters of the model are α, the ratio of the core+shell radius over the radius of the core i.e. α=R$_{core+shell}$/R$_{core}$ and $\epsilon_r$ the permittivity of the shell. The case α=1 would refer to the situation with no shell, i.e. the single crystal case; in other case we get α>1. Experimentally, the shell could be either a "natural" shell, for instance in the case of a BT ceramic, some defective BT at the surface of the grains with frozen polarization and low permittivity (in this case we use typically $\epsilon_r$≈75-100), or an "artificial" shell resulting from a coating of the BT grains, for instance SiO$_2$ ($\epsilon_r$≈5). Afterward in the modeling the ceramic is built on these core-shell particles, and one applies an external electric field, using the Lorentz equation of a particle embedded in a medium. In the mean field approximation, each particle sees the external field plus the mean field created by all the other particles:

$$E_{core,i} = \epsilon_0^{-1}\beta P_{core,i} + \gamma E_{ext,i} \qquad (1)$$

where i=x,y,z, $P_{core}$ corresponds to the polarization of the core, $E_{ext}$ is the external applied electric field, $\epsilon_0$ is the vacuum permittivity, and $\beta$ and $\gamma$ are calculated coefficients that are detailed below. The volume polarization P$_{core-shell}$ of the whole core-shell (and therefore of the ceramic) is equal to:

$$P_{core-shell,i} = \beta' P_{core,i} + \gamma' E_{ext,i} \quad (2)$$

The parameters $\beta$, $\beta'$, $\gamma$, $\gamma'$ have simple analytical forms[9] which are easily calculated from the 2 parameters of the model, α and $\epsilon_r$. One observes that inside the ferroelectric, it is no more $E_{ext}$ but $\gamma E_{ext}$ which is felt, and also the existence of a depolarizing field proportional to $\beta$, which is therefore an effective depolarizing coefficient. In the simple case of a spherical particles one gets[9]:

$$\beta = \frac{1-\alpha^3}{(2\alpha^3+1)\epsilon_r+\alpha^3-1} \qquad (3)$$

$$\gamma = \frac{3\alpha^3 \epsilon_r}{(2\alpha^3 + 1)\epsilon_r + \alpha^3 - 1}$$

$$\beta' = \frac{3\epsilon_r}{(2\alpha^3 + 1)\epsilon_r + \alpha^3 - 1}$$

$$\gamma' = \frac{(\alpha^3 - 1)(2\epsilon_0\epsilon_r^2 - \epsilon_0\epsilon_r - \epsilon_0)}{(2\alpha^3 + 1)\epsilon_r + \alpha^3 - 1}$$

This means that contrarily to an infinite system, a core-shell ceramic feels a microscopic depolarizing field. Note also that in the case of non-spherical particles the corresponding tensors $\beta_i$ etc have also been calculated[9]. When performing a Landau expansion of the free energy of the core on the components $P_{core,i}$ we therefore must add to the classical term:

$$F = \alpha_0(T - T_c) \sum_i P_{core,i}^2 + etc \dots \quad (4)$$

the 2 terms :

$$\frac{1}{2}\beta \sum_i \frac{P_{core,i}^2}{\epsilon_0} - \gamma \sum_i P_{core,i} E_{ext,i} \quad (5)$$

where the first added term correspond to the depolarizing energy and second term to the energy due to the external electric field. From now we make the assumption that the core of the grain is mono-domain and we consider that grains are orientated along one threefold [100] cubic direction; other directions have no qualitative impact on the results of the present study; we can then ignore the i-components.

III Effective permittivity of ideal core-shell ceramic

IIIa Case of a linear dielectric core

From the model developed above we can now deduce the effective permittivity $\epsilon_{eff}$ of the core-shell ceramic. Indeed we have for the polarization of the core:

$$P_{core} = \epsilon_0(\epsilon_i - 1)E_{core} \quad (6)$$

where $\epsilon_i$ is the permittivity of the core.

From equation (6) and equation (1), we get the self-consistent equation:

$$P_{core} = \epsilon_0(\epsilon_i - 1)(\epsilon_0^{-1}\beta P_{core} + \gamma E_{ext}) \quad (7)$$

Equation 7 allows to extract $P_{core}$, using the expression of $\beta$ and $\gamma$ from (3):

$$P_{core} = \frac{3\alpha^3 \epsilon_r \epsilon_0 (\epsilon_i - 1)}{(2\alpha^3 + 1)\epsilon_r + (\alpha^3 - 1)\epsilon_i} E_{ext} \quad (8)$$

For the polarization of the core-shell we have:

$$P_{core-shell} = \epsilon_0(\epsilon_{eff} - 1)E_{ext} = \beta' P_{core} + \gamma' E_{ext} \quad (9)$$

And thus finally using equation (8):

$$\epsilon_{eff} = \frac{(2\alpha^3-2)\epsilon_r^2 + (\alpha^3+2)\epsilon_r \epsilon_i}{(2\alpha^3+1)\epsilon_r + (\alpha^3-1)\epsilon_i} \quad (10)$$

We can notice that this formula is similar to the one obtained by Emelyanov et al.[14] in the case of the mixing of two dielectrics within an approximation where the core is supposed with high permittivity.

IIIb Case of a ferroelectric core

In fact as Emelyanov et al.[14] and latter Lin et al[15] noticed, formula 10 can be used whatever the high value for $\epsilon_i$ and in particular in the case of a ferroelectric material. Indeed the existence of a spontaneous polarization changes the numerical value of the permittivity but within the approximation of small electric field signal (which is the case for the experimental measurement of permittivity) it only adds a constant value to the locally linear variation that cancels in the derivation. Therefore we need now to calculate the permittivity of the core $\epsilon_i$. The Landau theory allows us to calculate via:

$$\epsilon_i - 1 = \frac{1}{\epsilon_0} \frac{dP_{core}}{dE_{core}} = \frac{1}{\epsilon_0} \frac{d^2 P_{core}}{d^2 F} \quad (11)$$

where $F$ is the free energy (4)+(5), with numerical coefficient of the Landau expansion from Ref. 15. If we then replace $\epsilon_i$ in the formula of the effective permittivity (equation (10)) by its numerical value calculated from equation (11) we get the temperature dependence of $\epsilon_{eff}$: we show on figure 3 the case of the cubic-tetragonal transition (for the "isotropic" component, of course the general calculation includes vectorial and tensorial forms in the equations) for different value of α. Notice that changing the value of α can be experimentally obtained either by changing the grains size, or by changing the shell thickness (or both). In particular the case of a pure core particles ceramic (α=1) is compared on figure 3 with different values of shell thickness (α>1). We observe that in addition to the shift toward low temperatures already observed in the critical temperature[9], a diminishing of the maximum value of permittivity and increase of diffusivity is observed: clearly the core-shell structure can explain the experimental observations obtained in most usual cases of ceramics.

IV Effective permittivity of real core-shell ceramic

We now assume that all grains in the ceramic have a g(r) Gaussian distribution of the core radius, but the shell thickness remains constant:

$$g(r) = \frac{1}{\sigma\sqrt{2\pi}} e^{\frac{-(r-\mu)^2}{2\sigma^2}} \quad (12)$$

We have performed numerical calculation for N grains (typical value of N≈1000) whose size distribution follows (12). For each grain we use the Landau calculation described above and at the end of the process the average permittivity is calculated from the implicit Bruggeman equation which is a mixing law originally developed for binary mixing of particles:

$$\frac{\epsilon_{eff}-1}{3\epsilon_{eff}} = (1-f)\frac{\epsilon_i-1}{2\epsilon_{eff}+\epsilon_i} + f\frac{\epsilon_e-1}{2\epsilon_{eff}+\epsilon_i} \qquad (13)$$

where $f$ is the volumic ratio of both phases with $\epsilon_i$ and $\epsilon_e$ for their respective permittivity.

We use in fact the generalized form of this equation:

$$\frac{\epsilon_{eff}-1}{3\epsilon_{eff}} = \sum_i^N g(r)\frac{\epsilon_i-1}{2\epsilon_{eff}+\epsilon_i} \qquad (14)$$

The results are shown on figure 4 for the case of a shell with 2nm thickness and different average values for core radius, with σ=100 and $\epsilon_r$=70. The three evolutions of figure 4 display the progressive transformation of a relatively well-defined permittivity with a high and sharp maximum at a definite temperature towards a situation where the diffusivity is very high and the maximum value is much lower than for sharp distribution of the grain sizes.

V Effective permittivity of IBLC (Maxwell-Wagner) core-shell ceramic

As mentioned in the introduction, observation of ultrahigh values of permittivity with weak thermal dependence have been observed in ferroelectric BT-based family as well as in CCTO family which doesn't have intrinsic ferroelectricity. In the first family two a-priory strange facts are: 1) the ferroelectric anomalies are still detected whereas the "flat" component is extremely high 2) as these observations have been performed in nanometric ceramics, huge shifts of the critical temperatures should have been observed, which is not so. Therefore the question of the real mechanism that is at play should be clarified or possible existence of several coupled mechanisms should be demonstrated.

V-I. IBLC effect in non-intrinsic ferroelectric compounds.

In order to explain the high permittivity induced by IBLC effect in non-intrinsic ferroelectric such as CCTO we apply the basic hypothesis of MW relaxation to a core-shell ceramic made of two linear (i.e. not ferroelectric) dielectrics, assuming that the core has now a finite value of conductivity $\sigma_i$ associated to the imaginary part of its permittivity $\epsilon_i$, neglecting the intrinsic loss, which is weak compared to the static loss at low frequency:

$$\epsilon_i = \epsilon_i' - i\epsilon_i'' \qquad (15)$$

with

$$\epsilon_i'' = -\frac{\sigma_i}{\varepsilon_0\omega} \qquad (16)$$

Thus now we can use equation (10), replacing $\epsilon_i$ by its complex form (15). This leads to:

$$\epsilon_{eff} = \frac{(2\alpha^3-2)\epsilon_r^2+(\alpha^3+2)\epsilon_r(\epsilon_i'+\frac{i\sigma}{\varepsilon_0\omega})}{(2\alpha^3+1)\epsilon_r+(\alpha^3-1)(\epsilon_i'+\frac{i\sigma}{\varepsilon_0\omega})} \qquad (17)$$

From this expression the MW behavior, which is (as pointed-out in Ref. 15) widely evidenced to exhibit a Debye-like dielectric behavior, is observed: namely, the dielectric constant exhibits a characteristic frequency dispersion kink accompanied by a peak in the corresponding dielectric loss and more importantly also an ultrahigh value in magnitude of permittivity: an example is plotted in Fig. 5, to be compared to the experimental values on Fig. 2a. At low frequency (quasi-static case ω -> 0 ) equation (17) simplifies in equation (18).

$$\epsilon_{eff} = \frac{(\alpha^3+2)\epsilon_r}{\alpha^3-1} \qquad (18)$$

From this equation we observe that for weak value of the shell thickness, i.e. α close to 1, the effective permittivity can be colossal. We have therefore reproduced the main feature of CCTO-family compounds from the core-shell model.

V-II IBLC effect in ferroelectrics compounds.

The case of <u>ferroelectric</u> IBLC can be treated in a similar way as in section V-I by assuming a core-shell structure in which now the ferroelectric core has a non-zero conductivity, being conductor or semiconductor. As usual the electric field in the core is given by equation (1) which becomes in the absence of external field:

$$E_{core} = \epsilon_0^{-1} \beta P_{core} \qquad (19)$$

However we cannot use for $\beta$ its value given by the first equation (3), because up to now we have neglected for the ferroelectric core the background permittivity $\epsilon_b$ which is no more negligible when conductivity exists and can be high. So in order to take into account $\epsilon_b$ we must add to equation (6) a permanent polarization (in addition to the spontaneous polarization) $P_b$ induced by the background permittivity. Then equation (6) transforms into:

$$P_{core} = \epsilon_0(\epsilon_i - 1)E_{core} + P_b \qquad (20)$$

which modifies the self-consistent equation (7) and gives immediately the new value for $\beta$:

$$\beta = -\frac{\alpha^3-1}{(2\alpha^3+1)\epsilon_r+(\alpha^3-1)\epsilon_b} \qquad (21)$$

When introducing conductivity, equation (21) becomes:

$$\beta = -\frac{\alpha^3-1}{(2\alpha^3+1)\epsilon_r+(\alpha^3-1)(\epsilon_i'+\frac{i\sigma}{\epsilon_0\omega})} \qquad (22)$$

From equation (22) we can observe that when the conductivity is high or at low frequency (quasi-static case) the ferroelectric core is no more submitted to a depolarizing field because of the screening by the conducting charges. Indeed the $\beta'$ coefficient have of course the same conductivity sigma dependence, and via equation 2 it can be directly seen that a finite conductivity modify the depolarizing field. This explains why in experimental observations many nano-grains are still tetragonal at room temperature, whereas when the grains are perfect insulating they are cubic due to the shifts of the critical temperature towards low temperatures induced by the core-shell structure.

Like the preceding section we need now to calculate the permittivity of the core $\epsilon_i$ from the Landau theory. As the Landau potential is formally a free energy in static field, it is justified, taking into account also the remarks about equation (22) to approximate the complex polarization, electric field and $\beta$ to their real part, as the imaginary part is ≈0 when ω≈0. The real part of $\beta$ is:

$$R(\beta) = -\frac{(\alpha^3-1)((2\alpha^3+1)\epsilon_r+(\alpha^3-1)\epsilon_b)}{((2\alpha^3+1)\epsilon_r+(\alpha^3-1)\epsilon_b)^2+\frac{(\alpha^3-1)^2\sigma^2}{\epsilon_0^2\omega^2}} \qquad (23)$$

Due to the conductivity this value becomes lower than in the case of the pure insulating case.

If we are interested in the permittivity at the cubic-tetragonal transition we may use the one-dimension free energy which gives up to the 8$^{th}$ order:

$$F = \sum_1^4 \alpha_{2i} P_{core}^{2i} - R(\beta)\frac{P_{core}^2}{\varepsilon_0} \qquad (24)$$

With $\alpha_2 \approx (T-T_c)$

Using equation (11) (in its tensorial forms) gives the values of the nm-components of susceptibility $\chi_{nm}^f$ (f stands for "ferroelectric") and thus of the permittivity $\epsilon_{nm}^f$. With these values we can now calculate the corresponding effective component from equation (17). The calculation is lengthy but straightforward, for instance for the diagonal components we get:

$$\epsilon_{nn} = \frac{(\epsilon_r^2(2K+\epsilon_r\epsilon_t(K+3))(\epsilon_r(2K+3)+\epsilon_t K)+\epsilon_i^2\epsilon_r K(K+3)}{(\epsilon_r(2K+3)+\epsilon_t K)^2+\epsilon_i^2 K^2} \qquad (25)$$

where $K = \alpha^3 - 1$

$\epsilon_r = \epsilon_b + \epsilon_{nm}^f$

$\epsilon_i = -\frac{\sigma}{\varepsilon_0\omega} \qquad (26)$

Equation (25) allows us to plot for instance the trace of principal components on figure 6 for different values of $\epsilon_i$ which parametrizes both conduction and Maxwell-Wagner frequency. It is clearly seen that when the ferroelectric core is less and less submitted to the depolarizing field its permittivity is progressively masked by the permittivity induced by conductivity. However at very high value of $\epsilon_i$ the ferroelectric anomaly is much weaker than the one observed experimentally (e.g. Fig. 1).

However a closer agreement of the model to the experimental reality can be easily obtained if we take into account the inhomogeneity of the ceramic, assuming that some grains have not (semi)conductor core but remain purely ferroelectric. This is just the case of a percolative composite. In this case we can again use the Bruggemann equation (13) now describing a mixing of particles of same size but with a f-fraction of grains remaining ferroelectric and 1-f being (semi)conducting:

$$\frac{\epsilon_{eff}-1}{3\epsilon_{eff}} = (1-f)\frac{\epsilon_c-1}{2\epsilon_{eff}+\epsilon_c} + f\frac{\epsilon_i-1}{2\epsilon_{eff}+\epsilon_i} \qquad (27)$$

where $\epsilon_i$ is as above the core permittivity of the ferroelectric grains, as in equation (25) but which can be approximated by equation (18) and where $\epsilon_c$ is the permittivity of the (semi)conductive grains given by equation (10). The self-consistent equation (27) is calculated numerically and examples are shown in figure 7. In this case contrarily to the case in figure 6 the anomaly of permittivity doesn't disappears even for very high value of conductivity.

A further step can be done by taking into account size distribution $g_i(r)$ as we did in section IV, using now.

$$\frac{\epsilon_{eff}-1}{3\epsilon_{eff}} = \sum_j^{Nnc} g_j(r)\frac{\epsilon_{ij}-1}{2\epsilon_{eff}+\epsilon_{ij}} + \sum_j^{Nc} g_j(r)\frac{\epsilon_{cj}-1}{2\epsilon_{eff}+\epsilon_{cj}} \qquad (28)$$

We now get temperature-dependences (Figure 8) closer to the experimental case of figure 1, increasing the magnitude of conductivity by a 10-factors i.e. $\epsilon_i = 10^6$ and putting α=1.0005: this may experimentally correspond to the case of a micronic ceramic ≈4 µm with a 2 nm shell.

In a real ceramic there is probably no abrupt change between grains, and a more realistic description should be given by taking into account a slight conductivity distribution, which we didn't do. Therefore a final refinement of the model should have been to introduce a Gaussian distribution for the conductivity in addition to the size distribution but we didn't perform the calculation as we think the agreement with experiments is already fair.

VI General considerations on porosity in ferroelectric ceramics

It is obvious that ceramics with the best performances should have a density as close to 100% as possible. That means that both open and closed porosity must be avoided. Whereas open porosity is (more or less) easily avoided, closed porosity is more difficult to be minimized. As soon as there is a departure from the 100% ideal value, corrections in the measurement of dielectric permittivity must be applied, either in the form of a rough proportional correction or with the more sophisticated correction using Rushman and Striven equation[17]. But these corrections are no more effective as soon as the density is less, say, than 80%.

In order to understand what are the consequences of closed porosity a simplistic model of ceramic built on pure core monodomain spherical grains with a $P_{core}$ polarization and a $f$ volume fraction of closed porosity is enough to get the main features. Then the total average polarization is:

$$P_{total} = (1-f)P_{core} \qquad (29)$$

Normally to calculate $P_{core}$ one should, as done several times above, consider the effective response of the surrounding media on the grain, but here we assume that $f$ is weak enough in such a way that the permittivity of the grain and of the total ceramic are almost the same. As the spherical grain is surrounded by other polarized spheres, we are in the classical case described by Lorentz of a spherical cavity in an effective polarized medium, in such a way that the electric field felt by one grain due to the surrounding medium is the Lorentz field, weighed by the amount of porosity $(1-f)\frac{P_{core}}{3\epsilon_0}$ to which we have to add the electric field created inside the grain, $-\frac{P_{core}}{3\epsilon_0}$ which gives:

$$E_{core} = -\frac{P_{core}}{3\epsilon_0} + (1-f)\frac{P_{core}}{3\epsilon_0} = -f\frac{P_{core}}{3\epsilon_0} \quad (30)$$

This electric field induces in the Landau potential a depolarizing energy term of the form:

$$\mathcal{E}_{dep} = -f\frac{P_{core}^2}{6\epsilon_0} \quad (31)$$

which straightforwardly induces a renormalization of the critical temperature with a shift:

$$\Delta T = \frac{f}{6\alpha_0\epsilon_0} \quad (32)$$

where $\alpha_0$ is the first order dielectric rigidity.

For BT $\alpha_0$ is 4.124 $10^5$ which means that for value of $f$ as low as 0.01 (1% of volumic porosity) we get a $\Delta T$ as high as ≈500K! This means that a BT ceramic should never be ferroelectric, which obviously is not the case. In order to reconcile the model with the experimental reality we must once again take into account the conductivity of the grains (Fig. 9). Due to this conductivity, charges move to the interfaces between the grains and the porosity. Therefore the "dielectric" effect of the porosity is partially or fully screened by these mobile charges.

This simple description shows that the porosity should be described just like a dielectric inclusion inside a perfect (on the standpoint of density) ceramic, both forming a Bruggeman composite: this explains why porosity has not a so strong influence on the existence of ferroelectricity, its effects are just dilution effects just like in a composite.

VII Practical use of an IBLC ferroelectric and comparison with insulating core-shell

Although IBLC high values of permittivity result from extrinsic phenomenon due to charges at interfaces, the interest to use such compounds for instance to make energy storage under static field has been proposed in literature. Indeed in the case of an insulating core, we have already shown that there is a great advantage to have a shell because the energy that can be stored in such a configuration can be multiplied by a factor of almost ten (se e.g. figure 10 of Ref. 8) for an optimal value of α. This is due to the fact that the existence of a shell shifts the saturation of polarization to higher electric field.

If we assume now that the ceramic is built on grains with conductor core and insulating shell this has the oblivious consequence that no energy can be store in the conductive part of the ceramic, which means that all the energy stored is close to surface, in the shell. In order to calculate its value we need to know the average value of the polarization in the shell $\langle P_{shell} \rangle$ as it has not a homogenous value. This can be done in principle by making integration of the electric field inside the shell, but as the calculation in spherical coordinates are fastidious we may use a simpler approximation and write, with the different volume $V_{core}$, $V_{shell}$ and $V_{core-shell}$:

$$\langle P_{shell} \rangle = \frac{P_{core-shell}V_{core-shell} - P_{core}V_{core}}{V_{shell}} \quad (33)$$

which becomes for a spherical particle:

$$\langle P_{shell} \rangle = \frac{\alpha^3 P_{core-shell} - P_{core}}{\alpha^3 - 1} \quad (34)$$

We need now to calculate $P_{core-shell}$ and $P_{core}$.

$P_{core}$ is obtained from equation (1) and putting $E_{core}$ equal to zero because now the core is conducting, therefore the electric field inside the core is completely screened at low frequencies. So this leads to:

$$P_{core} = -\frac{\gamma}{\beta} \epsilon_0 E_{ext} \quad (35)$$

where the ratio of γ and β is given from equations (3). It gives:

$$P_{core} = \epsilon_0 \epsilon_r \left(\frac{3\alpha^3}{\alpha^3 - 1}\right)^2 E_{ext} \quad (36)$$

$P_{core-shell}$ is straightforwardly obtained from equation (2), using the value of $P_{core}$ from equation (36) and values of γ' and β' from equations (3). The value of $P_{core-shell}$ is thus:

$$P_{core-shell} = \frac{\epsilon_0 [(\alpha^3 + 2)\epsilon_r + (1 - \alpha^3)]}{\alpha^3 - 1} E_{ext} \quad (37)$$

Inserting the value of $P_{core-shell}$ and $P_{core}$ given by equations (36) and (37) in (34) gives the value of the mean polarization in the shell:

$$\langle P_{shell} \rangle = \frac{\epsilon_0 \alpha^3 (\epsilon_r - 1)}{\alpha^3 - 1} E_{ext} \quad (38)$$

As $\langle P_{shell} \rangle = \epsilon_0 (\epsilon_r - 1) \langle E_{shell} \rangle \quad (39)$

where the average $\langle E_{shell} \rangle$ is non-zero only along the direction of the applied field, due to axial symmetry, the average along others directions been zero, this lead to:

$$\langle E_{shell} \rangle = \left(\frac{\alpha^3}{\alpha^3 - 1}\right) E_{ext} \quad (40)$$

We notice that, as expected, the electric field in the shell is fully redistributed in the shell thickness.

As we are interested to know if there is a potential interest to store electrical energy in the IBLC compounds we must calculate the energy density $\mathcal{E}_{shell}$ :

$$\mathcal{E}_{shell} = \frac{1}{2} \epsilon_0 \epsilon_r \langle E_{shell} \rangle^2 = \frac{1}{2} \epsilon_0 \epsilon_r \left(\frac{\alpha^3}{\alpha^3 - 1}\right)^2 E_{ext}^2 \quad (41)$$

As all this energy is stored in the shell, we obtain for the total energy density $\mathcal{E}_{total}$ :

$$\mathcal{E}_{total} = \left(\frac{\alpha^3 - 1}{\alpha^3}\right) \mathcal{E}_{shell} = \frac{1}{2} \epsilon_0 \epsilon_r \left(\frac{\alpha^3 - 1}{\alpha^3}\right) \left(\frac{\alpha^3}{\alpha^3 - 1}\right)^2 E_{ext}^2 = \frac{1}{2} \epsilon_0 \epsilon_r \left(\frac{\alpha^3}{\alpha^3 - 1}\right) E_{ext}^2 \quad (42)$$

We observe that the smallest the thickness of the shell, the highest the energy that can be stored. But of course there is the experimental limitation of breaking field $E_{brk}$. If we want to apply a maximal field up to $E_{brk}$ on the shell, we must apply on the ceramic a field:

$$E_{ext}^{max} = \left(\frac{\alpha^3-1}{\alpha^3}\right) E_{brk} \quad (43)$$

For instance with a SiO₂ shell ($\epsilon_r = 5$), it is know that at nanometric size, $E_{brk}$ can be very large up to ≈ 10³ MV/m. With a conductive BT with ≈200 nm grains size, coated by a shell of SiO₂ of ≈20 nm, i.e. relative thickness α≈ 1.10, equation (43) shows that we should apply a field of 240 10⁶ V.m⁻¹ which is a rather unrealistic value for standard applications. On the other hand, in practice much lower values of breakdown field are very often found due to non-intrinsic effects related to the quality of the ceramic (density, porosity, mechanical and chemical defects etc) and the limitation stands in-between the maximum voltage that can be used and the damages that can happen due to the electric field.

In Ref. 9 (Tab II) we have shown that for a non-conductive core-shell of BT@SiO2 , with a standard $E_{ext}$ = 10MV/m , the maximum energy density that can be store is 1.8 MJ/m³ with the optimal value of α ≈ 1.0030 , to be compared with 1.6 GJ/m³ for a Li-battery . Now in the case of a conductive BT, equation (43) gives the optimal value for α with the same $E_{ext}$ and $E_{brk}$ ≈ 10³ MV/m :

$$\alpha = \left(\frac{E_{brk}}{E_{brk}-E_{app}}\right)^{\frac{1}{3}} \approx 1.0033 \quad (44)$$

which is almost the same optimal value for the insulating BT@SiO2 (Ref. 9, Fig.10). However the density of energy, given by (42), is in this case 0.22MJ/m³, which is 10 time smaller than for the insulating core-shell. The lowering of the value is strong but not so strong that could have been though. Moreover if we consider now a more easy to be synthesized value of α = 1.01 (but far less optimal for energy storage) one passes from 0.13 MJ/m³ for the insulating case to 0.075 MJ/m³ i.e. a lowering less than a factor 2. The global conclusion of this section is that IBLC have some lower interest for energy storage than standard core-shell structure. However as the limit imposed by the intrinsic breakdown field is far from been reached, if improvement of the ceramic quality could be achieved, higher values of voltage could be reached, and a factor 2 or more could be achieved on the energy storage.

VIII Conclusion

In this paper we have use a core-shell model of ferroelectric ceramic, previously applied to energy storage and electrocaloric effect and modify it to study temperature evolution of high values of permittivity reported in the literature. The physical consequences of the existence of a shell, either due to defective surface or coating of the grains is a screening of polarization due to a depolarization field whose magnitude is a function of the relative thickness of the core and shell as well as of the permittivity of the shell. With this model, size-dependence of both amplitude and temperature of the ferroelectric anomalies are calculated with the assumption of a perfect insulating ferroelectric core and dielectric shell. Diminishing of amplitude, increasing of diffusivity and shift of temperature of the anomaly could be reproduced. Introducing a size distribution of grains is shown to be responsible of a further reduction of the maximum value as well as diffusivity of the anomalies.

We modified afterward the model to reproduce the experimental behavior of the CCTO-family and the IBLC BT-based compounds. For that purpose we introduced a conductivity of the dielectric or ferroelectric core and we showed that it induces a huge background of permittivity and a Maxwell Wagner relaxation. In these systems both electric field and associated energy are now distributed only in the shell. In the IBLC BT family a simple coexistence of both insulating and conductive grains was shown to explain the persistence of the ferroelectric anomaly even though a huge flat component of permittivity is observed. A fair agreement with the experimental situation was obtained when allowing a size distribution of the grain. Porosity was briefly considered and we observed that it can be considered just like inclusions in the ferroelectric matrix whose mobile charges screen the strong effects on the transition.

Finally we have evaluated the potentiality of energy storage in IBLC BT-compounds compared with standard (insulating) BT. We have observed a diminishing of a factor $\approx$4 for optimal core-shell structure, and of a factor $\approx$2 far from this optimum which shows that the performance of IBLC compounds are lower than optimal core-shell structures.

Examining the physical consequences of the different parameters at play in the dielectric properties discussed in this paper should help, we hope, to understand and tailor the ceramics in order to get adequate performances and to minimize the parasitic effects that affect them.

.

à

Figures captions :

Figure 1 : Permittivity at 10 kHz of a core@shell ceramic of $BaTiO_3$@$SiO_2$ particles sintered by spark plasma sintering after a first and second post-annealing (reprinted from Ref. 3).

Figure 2: a) Temperature-dependence of the real part of the dielectric response of $Cu_3Ti_4O_{12}$, at different frequencies between 100 (white circles) and 1 MHz (white triangles) (reprinted from Ref.5). b) Frequency-dependence (from Ref. 7).

Figure 3: Calculation of the temperature-dependence of permittivity of a perfect core-shell ceramic, for different radius ratio α=$R_{core+shell}$/$R_{core}$ ; α = 1 is therefore the case with no shell, i.e. single crystal case. A grain size of 2μm for instance (i.e. $R_{core+shell}$ = 1μm) with α = 1.01 corresponds to the case of a shell thickness equal to 10nm.

Figure 4: Calculation of the temperature-dependence of permittivity in the case of a 2nm shell-thickness, for different types of grain size distribution, showing the progressive diminishing of maximum and increase of diffusivity.

Figure 5: Calculation of the frequency dependence of real and imaginary permittivity for IBLC such as CCTO (see Fig.2 b for experimental values), calculated from equation (17) with α=1.001, $\epsilon_r$ =10, $\epsilon_i$=100, $\epsilon_0$=8.854e-12, $\sigma$ =0.001.

Figure 6: Calculation of the temperature-dependence of permittivity in the case of a core-shell ceramic with conductive core. The plot shows the variation with the conductivity and Maxwell-Wagner frequency for a ceramic with α = 1.001 and $\epsilon_r$ = 70.

Figure 7: Calculation of the temperature-dependence of permittivity in the case of a ceramic with mixing of core-shells with insulating and conductive core. The plot shows the variation with the conductivity and Maxwell-Wagner frequency for a ceramic with α = 1.001 and $\epsilon_r$ = 70. f is the volumic ratio of insulating grains inside the ceramic.

Figure 8: Same as figure 6 but now with a Gaussian grain size distribution with $\epsilon_i$ = $10^6$ and putting α=1.0005: this may experimentally correspond to the case of a micronic ceramic ≈4 μm with a 2 nm shell.

Figure 9: Effect of porosity screened by mobile charges inside the ceramic.

Figure 1

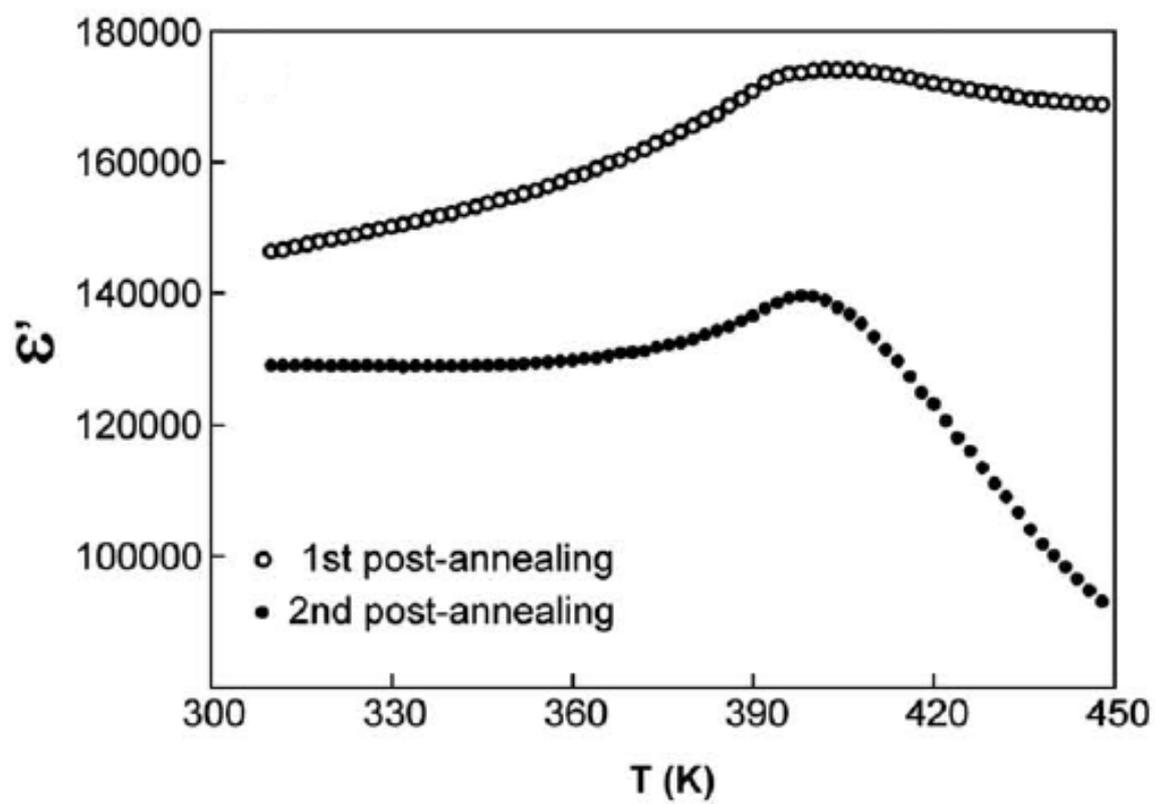

Figure 2

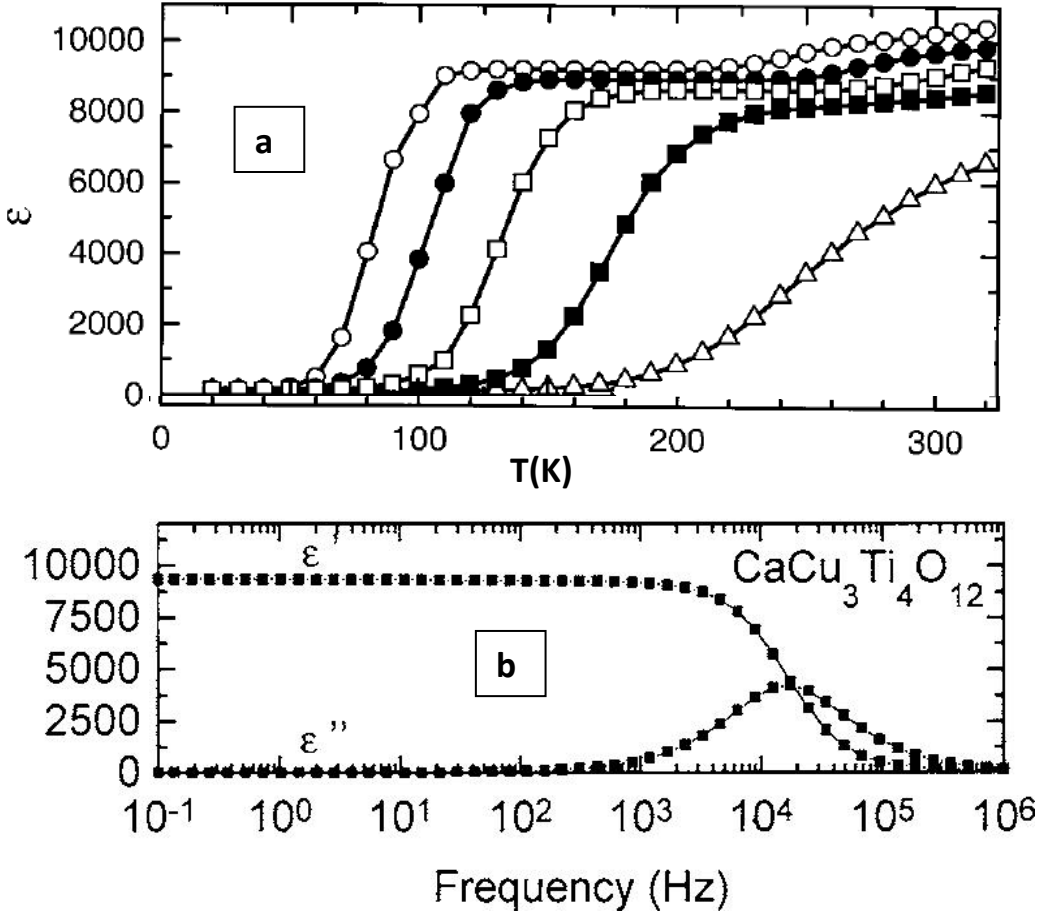

Figure 3

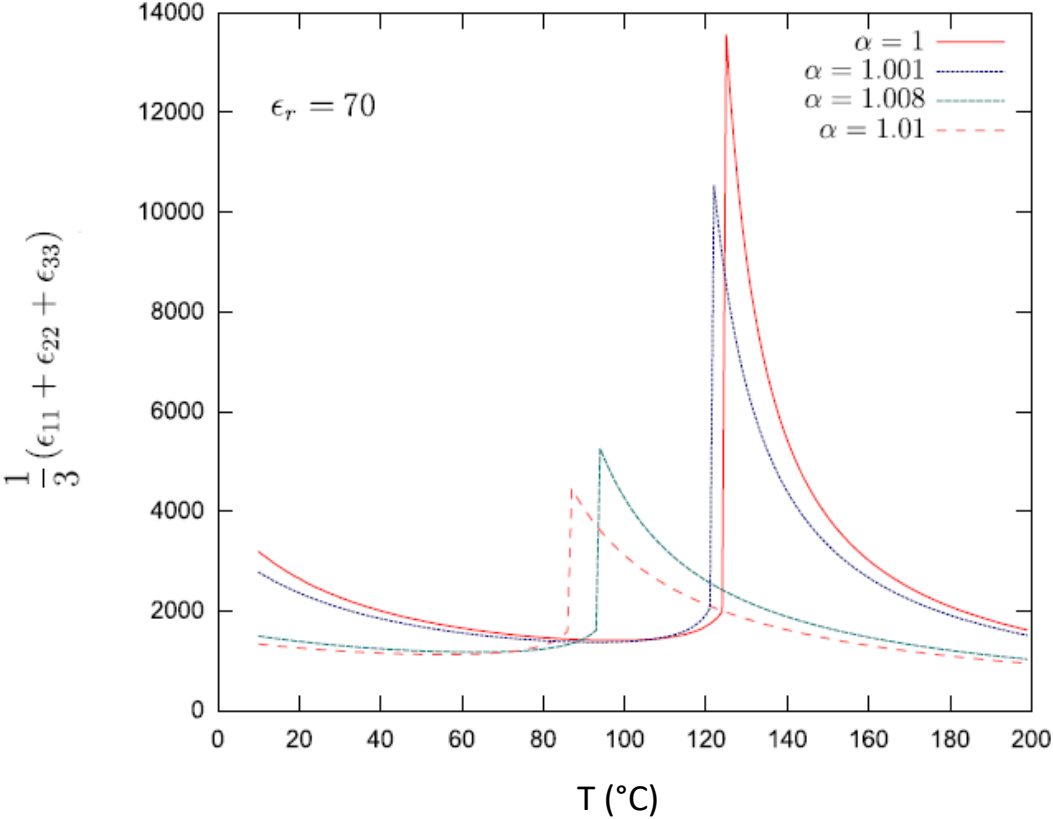

Figure 4

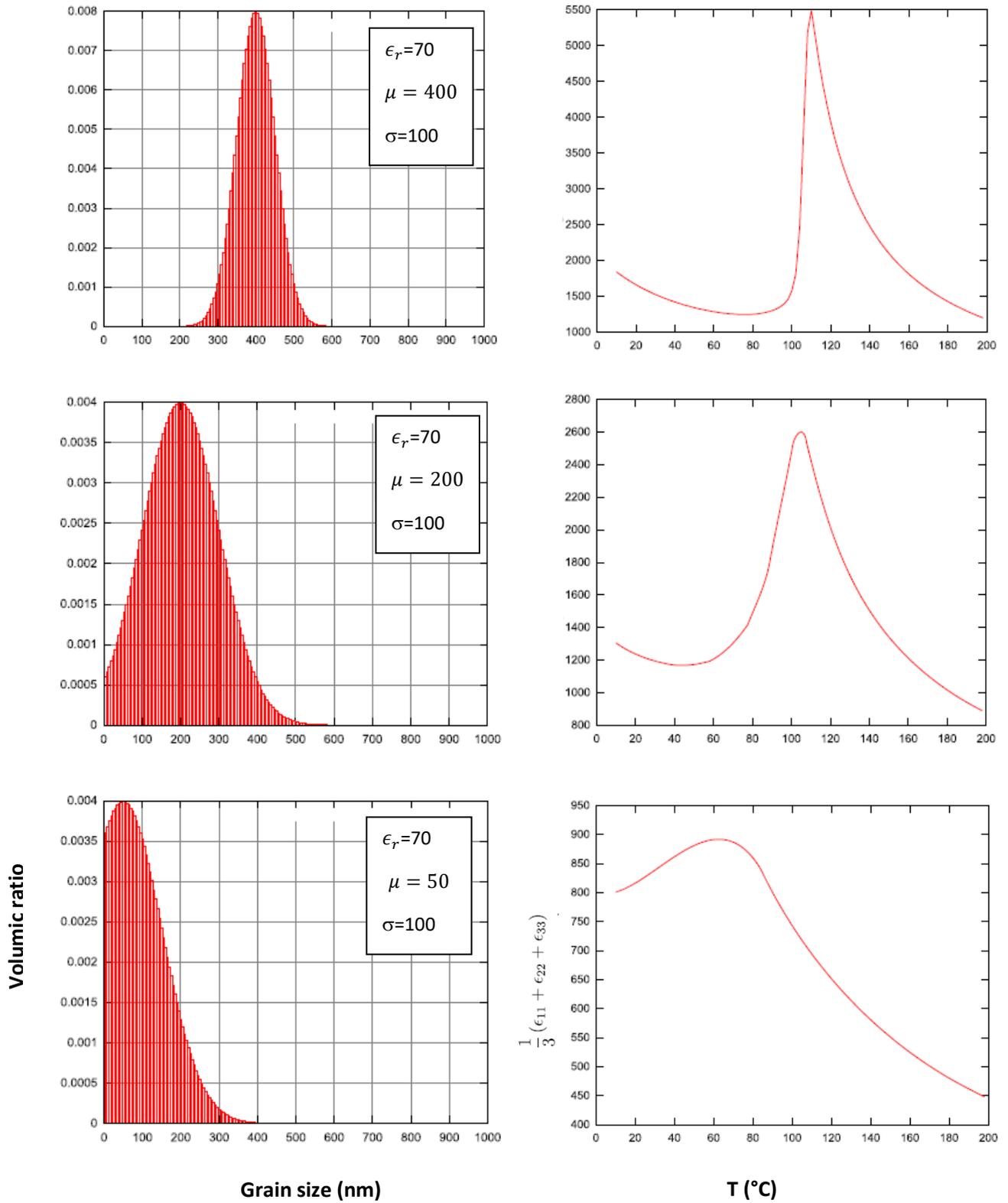

Figure 5

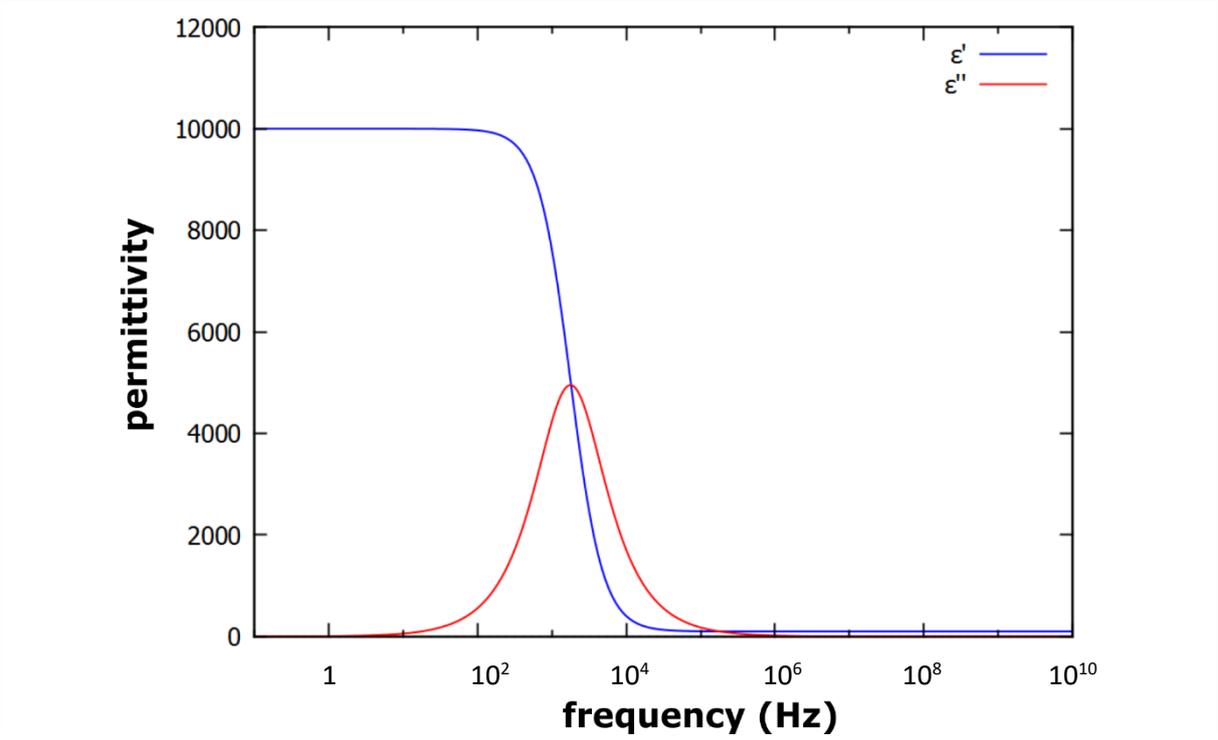

Figure 6

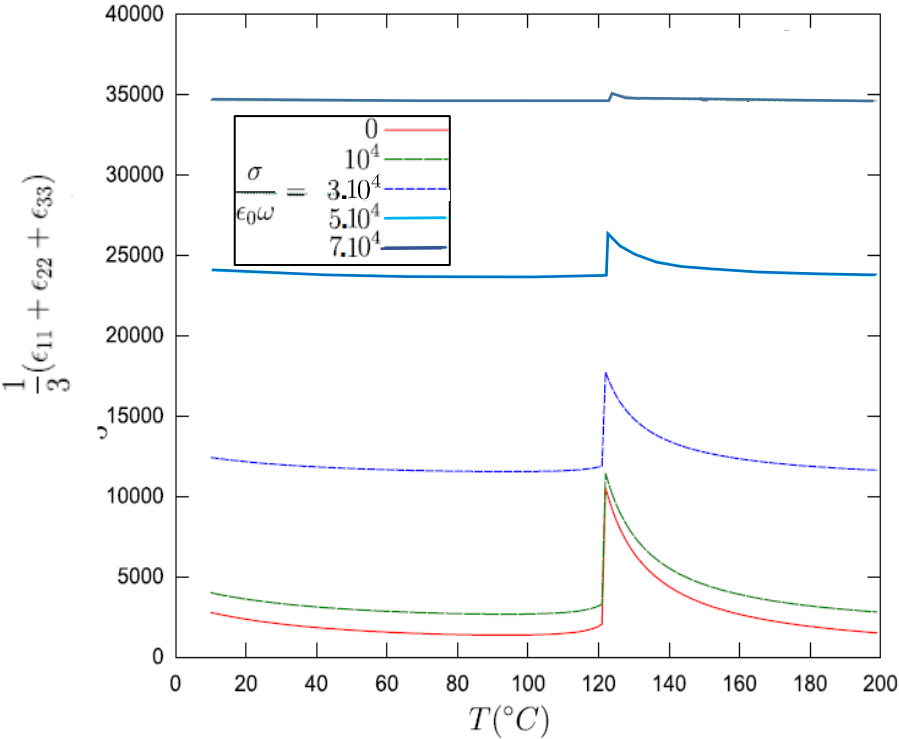

Figure 7

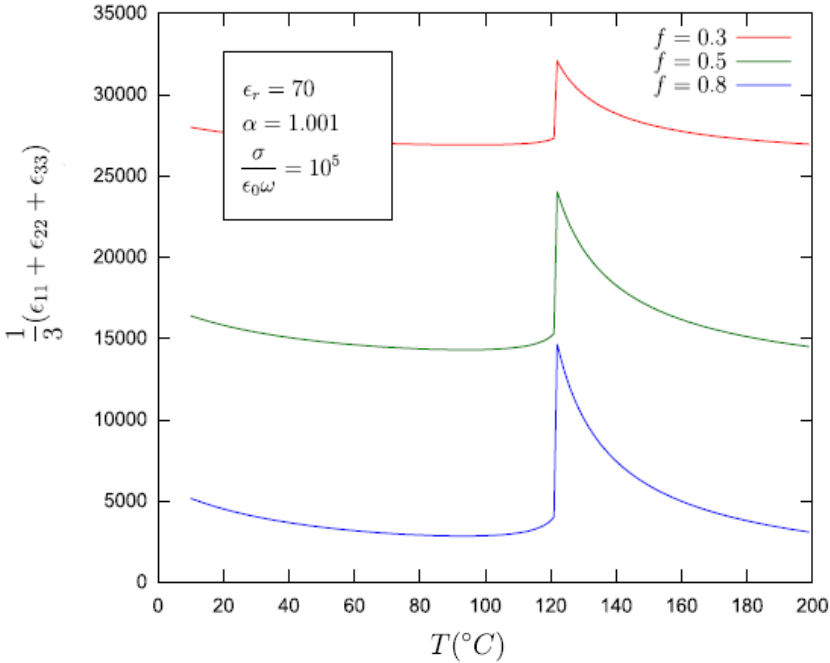

Figure 8

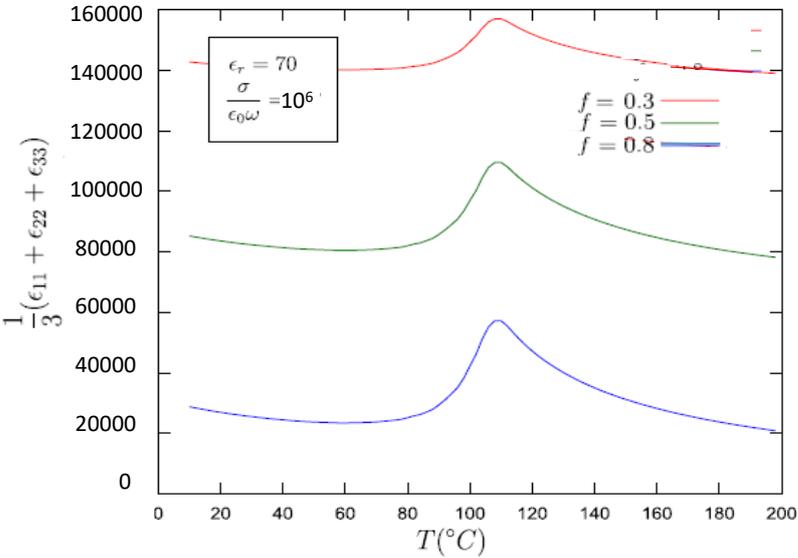

Figure 9

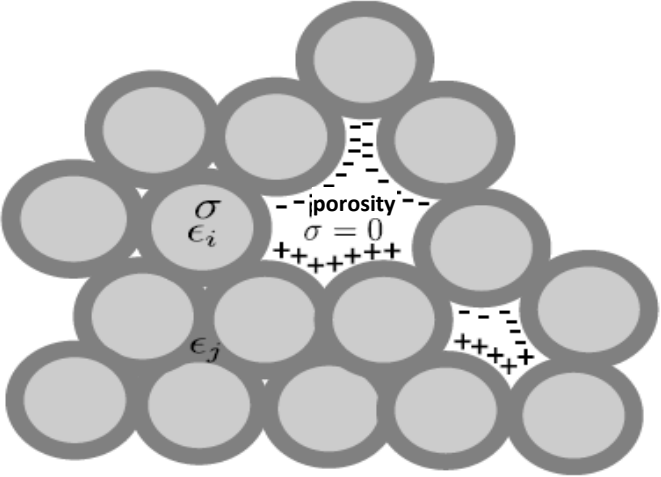